# Ethics Readiness of Artificial Intelligence:
# A Practical Evaluation Method

**Laurynas Adomaitis**,[1] **Vincent Israel-Jost**,[2] and **Alexei Grinbaum**[2]

**Abstract**

We present Ethics Readiness Levels (ERLs), a four-level, iterative method to track how ethical reflection is implemented in the design of AI systems. ERLs bridge high-level ethical principles and everyday engineering by turning ethical values into concrete prompts, checks, and controls within real use cases. The evaluation is conducted using a dynamic, tree-like questionnaire built from context-specific indicators, ensuring relevance to the technology and application domain. Beyond being a managerial tool, ERLs help facilitate a structured dialogue between ethics experts and technical teams, while our scoring system helps track progress over time. We demonstrate the methodology through two case studies: an AI facial sketch generator for law enforcement and a collaborative industrial robot. The ERL tool effectively catalyzes concrete design changes and promotes a shift from narrow technological solutionism to a more reflective, ethics-by-design mindset.

1. **Introduction**

We propose the idea of Ethics Readiness Levels (ERLs) as a structured tool for recurrently evaluating the integration of ethical thinking in research and software design processes. In the governance of emerging technologies, ethical guidance has often relied on so-called soft law instruments—codes of conduct, guidelines, or frameworks—designed to promote responsible behavior without imposing binding legal constraints. This is partly due to the difficulty of imposing harmonized regulations across the EU, especially in a global context characterized by strong reservations expressed by other international actors, e.g. the United States of America, with regard to the regulation of artificial intelligence (AI) that "unduly burdens AI innovation" (Kratsios, Sacks, and Rubio 2025). Another reason is related to the principle, upheld in several member states such as Germany, that protects scientific freedom by constitutional law. Nevertheless, the recent trajectory of technological regulation in the European Union shows that soft law can evolve into hard law: this has been the case, notably, with the adoption of the AI Act (European Commission 2022; Terpan 2015). Rather than opposing soft and hard law, our concern lies in their practical adaptation: how

---

[1] RISE Research Institutes of Sweden, Computer Science Department
[2] CEA-Saclay/Larsim

E-mail: laurynas.adomaitis@ri.se, vincent.israel.jost@cea.fr, alexei.grinbaum@cea.fr

can high-level societal values and principles, whether embedded in soft guidelines or hard regulation, be translated into operational steps for AI researchers and software developers?

One of the most influential soft law frameworks in Europe is the approach of 'Responsible Research and Innovation' (RRI), promoted to foster "acceptability, sustainability, and societal desirability" in scientific and technological innovation (Von Schomberg 2013). RRI signaled a new phase of institutional awareness, acknowledging the importance of societal concerns and ethical reflection throughout the innovation chain. It laid the groundwork for methodologies such as ethics-by-design and values-by-design (Brey and Dainow 2021; Van den Hoven, Vermaas, and Van de Poel 2015), which seek to embed ethical thinking directly into the development process. From a philosophical standpoint, the ethical approach championed in RRI is essentially deontological: it aims at codifying desirable conduct through normative principles, enabling ethical judgments based on compliance. This brings it close to the logic of legal and regulatory oversight. For example, the high-level principles and values listed in recital 27 of the EU AI Act are translated into a set of technical norms and standards requiring legal compliance.

Yet this deontological approach to AI regulation has long faced criticism. It tends to reduce responsibility to preordained norms, which may not capture the evolving and context-dependent nature of responsibility (Alexei Grinbaum and Groves 2013). Furthermore, these principles require a level of operationalization beyond usual legal parlance, making them technically meaningful and actionable for AI practitioners. For example, the principle of "respect for human rights" cannot simply remain a statement in natural language. As Lessig famously put it, "code is law" (Lessig 2000) if a principle cannot be translated into design decisions or constraints on the behavior of AI systems, it lacks operational effect. What exactly "respect for human rights" means in an operational setting is highly context-dependent: which humans, which rights, which criteria of respect cannot be decided via a preordained norm. The emergence of large language models (LLMs) adds a new twist to this well-known observation of content-dependence in technology ethics: AI systems seem to understand and respond to normative instructions expressed in natural language, giving the impression that principles such as "respect for human rights" could be operationalized simply as a consequence of being fed to the AI system in a prompt. This illusion has led to the development of approaches like "constitutional AI" (Kundu et al. 2023), in which LLMs are fine-tuned with high-level ethical directives. It creates a misleading sense of moral adequacy: the model only creates an illusion of understanding in the user, nor does it reason morally. It merely reproduces linguistic associations that correlate with certain user expectations, while robustness is lacking. As a result, users may believe for some time that the system is behaving based on rules—until an unexpected output breaks that illusion and reveals the absence of genuine ethical grounding (A. Grinbaum 2019). Rather than resolving the problem of operationalization, LLMs risk masking it.

The AI ethics community has increasingly turned toward more structured and transparent forms of operationalizing high-level ethical principles. A central example is the ALTAI checklist (European Commission 2020), as well as the GPAI code of practice. Both documents translate abstract values into a set of design questions addressed to software developers. These checklists—sometimes tailored to a specific domain, system type, or ethical concern—help guide decision-making, even if their usefulness remains contested. Critics warn that checklists can encourage a superficial "box-ticking" mentality (Stahl et al. 2019; Macnaghten 2020), be too long to be practical or too generic to remain meaningful in context (Mertz 2019; Nordtug & Haldar 2024).

Despite these valid criticisms, checklists can play a valuable role. Most stakeholders in the innovation chain are not professional ethicists, yet their management processes need to make room for ethical considerations, because the AI systems they build produce a major societal impact. Operational ethics management tools can help them integrate ethical thinking into their workflow, raise awareness, and guide reflection systematically (Politi and Grinbaum 2020). We propose to make use of checklists, not as static tools with a mere box-ticking approach, but rather as a dynamic and adaptive way of framing ethical reflection. Our tool facilitates this process and provides a dynamic and comparative evaluation of ethical thinking in software design.

1. **Ethics Readiness Evaluation methodology**
    1. **Background of readiness levels**

NASA developed TRLs in the 1970s assessing the maturity of technologies for space programs to avoid failed projects, overrun budgets, and mismatched expectations (Olechowski, Eppinger, and Joglekar 2015). Similarly, we present here an iterative process to track ethics readiness over time and design iterations to provide measurable answers to the questions of ethics readiness of AI systems during their design. Is the design process enabling or losing control of ethical issues over time? Are critical issues addressed before or after deployment? Are ethics concerns raised at critical stages during development for them to be adequately addressed? The ethics readiness tool presented here is meant as a measurement for managing this process and for avoiding unforeseen and uncontrolled impacts in sensitive areas when the system is deployed.

The evaluation methodology and tool are based on the idea of Ethics Readiness Levels (ERL) that use indicators to capture the state of readiness during the development of an AI system.

AI raises concerns that are not limited to previously known and regulated issues in safety and privacy. Some examples of novel issues include human autonomy, fairness, responsibility, and others. Several frameworks have been created for legal readiness levels specific to machine learning (Lavin et al. 2022) and blockchain technologies (Tang et al. 2020), however they do not address ethics concerns that go beyond compliance. There is a machine learning assessment tool that considers ethics evaluation as part of the TRL structure (Eljasik-Swoboda, Rathgeber, and Hasenauer 2019) but it presents no operational way to conduct such an evaluation. More recently, frameworks for the ethics readiness of technology have also emerged (Jong 2025), however they do not present a novel evaluation scale or evaluation methodology. Jong is the only recent source that presents ethics readiness levels, but they are matched to TRLs, so that it seems to be more of a feature of TRLs than an independent scale.

In what follows we present the details of a specific readiness scale for ethics (Adomaitis, Hoog, and Grinbaum 2024) and a novel methodology to conduct an ethics readiness evaluation. Lastly, we discuss two examples of mock evaluations to illustrate practical barriers and best practices.

## 2. Ethics readiness levels

To develop the level structure for ERLs, we loosely followed the integration readiness levels (IRLs) described by Sauser et al. (Sauser et al. 2006). IRLs reflect the reality that technologies do not exist independently but are connected through interfaces in the system architecture. A key aspect of the integration process is the identification-characterization-harmonisation-control paradigm. The four-level structure of ERLs aligns with the fundamental ideas in integration but transfers it into the domain of AI ethics of identifying, characterizing, harmonizing, and taking control of ethics issues.

ERL 0 indicates that development lacks ethical awareness. There is an absence of explicit ethical considerations. The AI system is presented as a neutral tool or solution, devoid of ethical implications.

ERL 1 encompasses the crucial step of identifying potential ethical dilemmas, privacy concerns, and legal implications associated with the system. This stage aligns with the ethical analysis process in technology ethics, which involves exploring the potential impacts of technology on stakeholders, dual-use scenarios, and the broader societal context (Adomaitis, Grinbaum, and Lenzi 2022; Brey 2010). The process entails a thorough review of the system, leveraging ethical theories and principles to pinpoint potential ethical issues (Umbrello et al. 2023). These identified issues serve as the foundation for subsequent stages of ethical reflection.

At ERL 2, the relationships and tradeoffs between various ethical issues are reflected. This stage makes use of ethical dilemmas in technology ethics, where ethical issues lead to conceptual conflicts or synergies (Stahl, Timmermans, and Flick 2017). The primary challenges at ERL 2 include the inherent uncertainty of the future (Dupuy and Grinbaum 2005) and the malleability of specific technologies, particularly AI (Moor 1985). Reflecting on the different interactions and trade-offs is needed for making informed decisions regarding how to effectively address the ethical issues identified at ERL 1.

ERL 3 shows operationalization and involves the integration of ethics considerations into the system design, a process often referred to as "ethics by design". This stage corresponds to the proactive approach to technology ethics, where ethical considerations are not reactive responses to ethical issues arising during application but are integrated into the design process (Van den Hoven, Vermaas, and Van de Poel 2015). This approach ensures that ethics is not an afterthought but an integral part of the system's design with material outputs.

Finally, ERL 4 shows control mechanisms for accountability and the attainment of standard benchmarks and certifications, if applicable. This stage reflects the importance of accountability in technology ethics (Floridi 2016; Jonas 1985), which not only ensures adherence and control of ethical issues but also enhances the system's credibility and trustworthiness, leading to trust among stakeholders and the broader public.

This ERL structure combines anticipatory ethics, where ethical considerations are anticipated before deploying the technology (Guston 2014), with ethics by design, that involves ethics as an active part of the development process. It allows management of emerging AI technologies while such management is still possible.

| Readiness Level Comparison | |
|---|---|
| *Ethics readiness levels* | *Corresponding IRL* (Sauser et al.) |
| ERL 0 – Ethics considerations lacking. | IRL 0 - Integration lacking |
| ERL 1 – Identified Ethics Issues: Ethics considerations raised by the system have been identified and anticipated. | IRL 1 - Interface identified w/ detail to allow characterization |
| ERL 2 – Characterised Interactions of Ethics issues: The interactions between different legal, ethical, and privacy considerations have been characterised. | IRL 2 - Specificity to characterize interaction between technologies |
| ERL 3 – Ethical Tensions Addressed via Ethics by Design: The system's ethical considerations have been designed to be compatible with each other and proactively implemented in the design of the system. Improving one aspect (e.g., system security) does not negatively impact another aspect (e.g., user accessibility), or these impacts have been optimized. | IRL 3 - Compatibility (common language) between technologies |
| ERL 4 – Control Over Ethics Issues: The system has implemented control mechanisms for accountability and has passed standard benchmarks and obtained certification, if applicable. | IRL 5 - Sufficient control between technologies to manage integration |

Table 1. A summary of ERLs with corresponding IRLs

## 2. Evaluation methodology for Ethics Readiness Levels

All steps in the ERLs – identification, characterization, harmonization, and control require placing ethical reflection in the operational contexts of a particular use case. The high-level principles and values need to become practically applicable in specific contexts. For example, transparency in financial fraud detection systems is not the same concept as transparency in agricultural robot navigation systems. They have something conceptual in common by invoking the thick concept of transparency. Concepts like 'fairness', 'transparency', 'privacy', and 'accountability' are "thick" because they are not purely descriptive, nor purely normative. They combine both dimensions. To call a system 'unfair' is both to describe a certain state of affairs (e.g., a disparate distribution of outcomes) and to make a negative judgment about it (Williams 2006).

For this reason, any viable ethics readiness tool that seeks to achieve actual impact on the implementation, must focus on the operational details in addition to discussing thick concepts. This creates a need for a dynamic and domain-specific questionnaire instead of a generic checklist. Domain-specific indicators provide a much more accurate representation of ethics readiness because they adapt to each use case being evaluated. In our methodology, this dynamic adaptation is achieved in two steps - indicator blocks and indicator structure.

Indicator blocks are the coarse-grained solution to adapting the evaluation based on the context of a use case. Blocks are sets of indicators that are grouped together based on a technology or the application area. Blocks can be included or excluded based on some preliminary information about the use case or the component being evaluated.

For example, a block of indicators can include the fact that the system is being designed to be used by law enforcement agencies (LEAs), a tool that uses personal data in their development, a tool that uses a specific type of AI system, like a general-purpose AI system (GPAI), etc. The intention is to tailor the evaluation to the use case by including and excluding certain blocks based on the use case. The fact that a certain system is meant to be only used by law enforcement can impact how ethics considerations are weighed. Thus, the blocks can depend both upon the type of technology, and on the area of application.

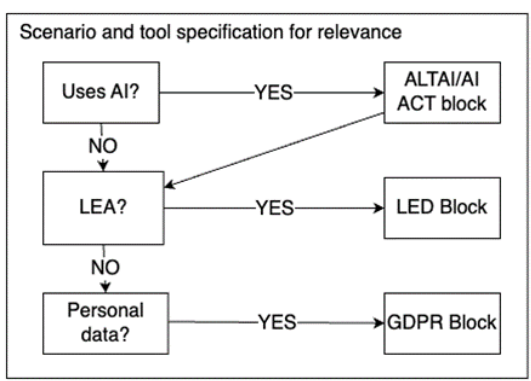

In the area of security broadly construed, the ERL blocks can include 1) a basic technology ethics module that all use cases are subject to, 2) AI-specific block, 3) LEA block, 4) Personal data block. It is important to note that depending on specific areas, blocks may be interrelated.

For example, if a use case is intended for LEA use, then they fall under the exception of GDPR and instead should respond to the Law Enforcement Directive (LED), regardless of the fact that it uses personal data. This creates a distinct legal reality for data protection in policing and judicial cooperation (de Hert and Papakonstantinou 2016). However, the EU AI Act does not provide a blanket exemption for AI systems simply because they are used by LEAs, meaning its requirements for high-risk systems apply.

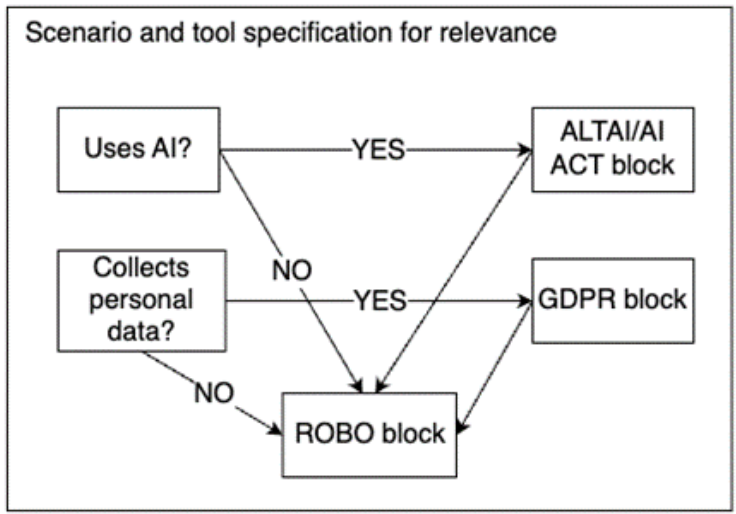

Indicator blocks may align very differently in robotics, where the GDPR and AI act apply together. The unique challenges of robotics, such as physical embodiment, human-robot interaction, and autonomous navigation, necessitate a tailored ethical evaluation beyond standard digital ethics checklists (Calo 2015; Alexei Grinbaum et al. 2017). Instead of having a generalist evaluation as the basis for all use cases, it should be adapted specifically to robotics, even though no general regulation exists.[3] This ensures that ethics readiness indicators specific to robotics are always addressed, and additional questions are only raised if applicable. This is a process of specification, where abstract norms are made concrete for a particular situation (Richardson 1990).

Each block contains a multitude of indicators. At the maximum, when all applicable blocks are included, our instance of the ERL tool included around 200 indicators. This number is lower if not all blocks are applicable. However, the number of indicators should be kept to a minimum to reduce the time needed to complete the ERL evaluation. Time is a critical constraint when implementing evaluation tools.

To reduce the time needed, and the amount of irrelevant indicators during the evaluation, we adopt a tree-like structure that only suggests further indicators if the answer to the lower-level indicator is "yes". If the answer is "no", the questionnaire passes on to the same order of indicator. If there is no same order indicator in the questionnaire, the next lower-level indicator is presented. When the next lower-level indicator does not exist, the questionnaire is over.

---

[3] These exemplary blocks are related to regulation only because regulation reflects sets of norms, but the evaluation should in no way be legalistic. The alignment of these sets of norms and the blocks is not necessary for the functioning of the tool, and an alternative division is equally viable.

This tree-like structure has significant implications for the whole evaluation, because it implies that not all use cases will be measured by the same indicators. This makes the evaluation very contextual and individual but renders the comparison among use cases very difficult. In different evaluation methodologies, there is a tension between standardization and individual treatment by use cases. Although compliance requires standardized treatment – that all use cases are treated the same way – ethics is much better conducted on an individual basis. Context-awareness is missing from broad ethics evaluations and standardization efforts, which has become one of the key points of criticism for them (Hagendorff 2020). ERLs methodology addresses this concern.

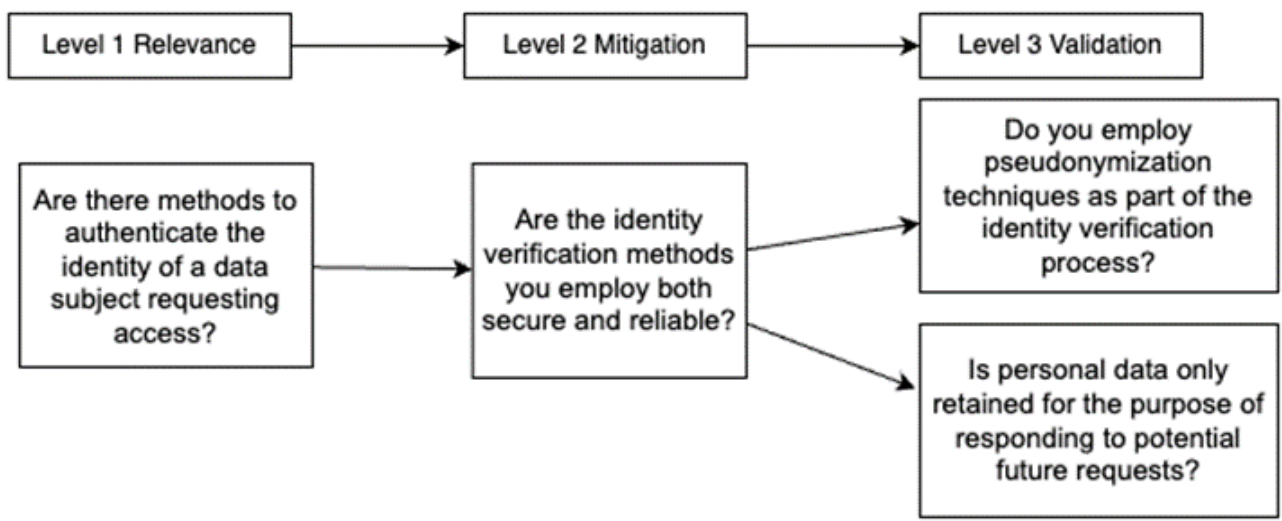

The tree-like structure only works if the semantics of the indicators allows for it. Indicators need to be formulated in a way in which they rely on each other. Typically, the lowest level indicator determines relevance - does this issue apply to the use case? If the answer is "no", the ERL tool skips all follow ups on that issue. If the answer is "yes", it moves to mitigation measures (middle layer). The middle layer determines whether mitigation was implemented to address the issue in question. Again, if the answer is "yes", the indicators branch out into validation indicators - are the mitigation techniques sufficient and well-implemented? The validation indicators help to address ethics-washing and validate that the mitigation is adequate and efficient (Bietti 2021).

These three layers are not always preserved, because there may be issues and sub-issues, or several steps in mitigating an issue, but generally indicators should preserve this relevance-mitigation-validation schemata because that is how ERLs are structured, based on the identification of issues, their conceptualization, harmonization, and control over them. Thus, the implicit structure of the indicators mimics the overall structure of the ERLs.

This adaptive questioning process allows the system to tailor the questionnaire dynamically based on the user's responses. It ensures that only relevant questions are presented to the user, and unnecessary questions are skipped. By following the tree-like structure of the indicator questions and considering the user's previous answers, the ERL tool efficiently navigates through the questionnaire, providing a personalised and streamlined user experience.

The sources for indicators in the areas of application can include existing guidelines, frameworks, legislation, or might need to be created from scratch. For example, there is no existing European legislation of robotics, although there is one for the use of technology by LEAs. Even if there is legislation, the language of it must be adapted to be understood as much as possible by the stakeholders involved in the ERL evaluation. Mostly, that includes developers and engineers.

The ERL tool is not intended for compliance or legal counselling, so it does not need to stick to the letter of the law. Instead, legal documents and guidelines are good representations of norms that can be used in the evaluation, but they should be used in a formulation that is understood easily and has an operational meaning.

For example, in practice, formulations of the indicators like 1) "Did you foresee any kind of external guidance or third-party auditing processes to oversee ethical concerns and accountability measures?" should be replaced with something like 2) "Can the product be audited by independent third parties?". It captures the same intention without involving legal notions that require explanation. In fact, the ERL tool should avoid the use of definitions and explanations because they detract from the dialogue, take a lot of time, and may require further definitions, or become circular.

Moreover, an important detail to consider is that the indicators only take "yes" and "no" for an answer. This design is useful for maintaining the tree-like structure of the indicators, because only by answering "yes" the user is presented with a follow up, but also saves time, and improves commitment to the answer. It is important to raise awareness and reflection on the question of what counts and satisfies an indicator.

3. **Scoring indicators and tracking progress over time**

Each answer to an indicator is scored and summed up to arrive at the final ethics readiness level. To indicate one of the four ERLs, the final score generally falls between 0 and 4.[4] To facilitate that, we postulate an initial score of 4 for all use cases. Thus, every evaluation session begins with a perfect score. Then, the logic of identifying, characterizing, and controlling relevant issues is implemented in the scoring. Usually, and depending on the formulation of each indicator, the lowest level indicator determines relevancy, and if the issue within the indicator applies, a negative score is added. In this example, the risk of significant damage means a score of -1, and the questionnaire continues with the mitigation measures of such damage. The scoring should be set up in such a way that if the use case implements all the measures, the total score under the root indicator adds up to 0. In this example, it is possible to regain a whole level back if all mitigation measures are implemented. However, each mitigation measure is again tested for adequacy with control indicators, and it may make the mitigation measure lose the positive score again.

---

[4] There may be negative scores, but they amount to ERL 0 since it still shows the lack of ethical reflection.

| number | indicator | yes_score | no_score |
|---|---|---|---|
| 2 | Can the system be manipulated to produce significant damage? | -1.000 | 0.000 |
| 2.1 | Does the product comply with recognized cybersecurity standards? | 0.360 | 0.000 |
| 2.2. | Are there additional security measures against potential attacks throughout the product's lifecycle? | 0.280 | 0.000 |
| 2.2.1 | Have you conducted penetration tests or red-team exercises on the product? | 0.000 | -0.280 |
| 2.3. | Does the timeframe for security updates match the expected lifespan of the product? | 0.080 | 0.000 |
| 2.3.1 | Have you informed users about the duration of security update coverage? | 0.000 | -0.080 |
| 2.4 | Can the product be misused for malicious or illegal purposes? | -0.270 | 0.000 |

| 2.4.1 | Have you conducted a risk assessment to identify potential misuse scenarios? | 0.180 | 0.000 |
| 2.4.1.1 | Have you implemented preventive measures to mitigate the risks of misuse? | 0.000 | -0.180 |
| 2.4.2 | Can users or third parties report misuse to the product designer? | 0.090 | 0.000 |

The scoring method deviates from the traditional indicators of TRLs. Usually, readiness indicators are taken to be necessary conditions for moving to the next level. For example, one would need to implement all measures before a positive score would be assigned. This can lead to the product being stuck at a certain readiness level, although other aspects of it may be more advanced.

In ethics, maturity cannot be defined in necessary conditions due to the opacity of ethics. There is no single and general predefined goal for the responsible development or use of technology. Such goals exist for TRLs, for example validating a prototype in a lab, or deploying a system operationally. Ethics includes an opaque element that depends on human desires, preferences, and changing norms. Indicators aim to capture all these elements, but there should be enough freedom to prioritize and make decisions based on the ethical reflection to provide a goal for each use case individually and the necessary conditions to achieve it. Therefore, the ERL tool seeks to capture only the direction of ethics readiness over time. The main question it tries to answer is whether a component or a use case is increasing or decreasing in ethics readiness, which is defined and refined through dialogue (Buber 2023; Levinas 1991). We discourage the use of this tool as self-assessment, which lacks dialogue, and comparison across applications or domains, since the evaluation always depends on the details of the use case.

The particular weights are assigned to each indicator via a two-stage approach. In the first stage, an intuitive list is made based on the consensus within the team that is developing the indicators. The issue with this method, of course, is that it is highly subjective. However, it provides an intuitive baseline and is better than random distribution of the scores. Generally, the biggest negative weights represent the highest posed risk (e.g. cybersecurity for critical infrastructure), or something that the applications in the area are very prone to (e.g. bias in public administration). Each ERL iteration in different domains should have an intuitive baseline so that it may be validated on real-life use cases.

Once a baseline of weights is established, validation exercises must be employed to make sure the weights are adequate. During this phase, the goal is to use the tool with engineers on real use cases and determine the clarity of the indicators (are they relevant, understandable, and spark reflection?), and the adequacy of the scores (are they too harsh or too lenient?). There are various methods and standards for validating such procedures but direct engagement with stakeholders and concrete feedback on the indicators and weights cannot be replaced. The size of the validation set of use cases should be around 10 use cases in the same area of application, although the validation process never ends and the weights can be adjusted with more exposure to real-life applications. If during the use of the ERL tool specific indicators or weights appear to be unclear or inadequate, they should be changed or reformulated.

4. **Results of the ethics readiness evaluation**

The evaluation sessions over time can be logged, and answers and scores can be collected for later comparison and reporting. It is also useful to visualize the scores, so that they produce a graph. This is an easy way to visualize which indicators produced the most negative or positive effect on the final score. By logging the sessions, answers, and scores, it is possible to track progress and identify trends over time. This information can be used to identify which indicators have the most significant impact on the final score, both positively and negatively. For example, if a particular indicator persistently receives negative scores, it may indicate an area that needs additional attention and reflection.

The frequency of the evaluation sessions depends on the resources available, but it is meaningful to follow up every 6 months and plan the sessions to be around 40 minutes long. The scale at which the evaluation is targeted depends on the goal of the evaluation. For example, in complex robotics systems it makes sense to evaluate different components (e.g. computer vision, navigation, task planning) separately, whereas a single recommender system may be better viewed as a whole, or as a use case. The ERL methodology can be adapted to work at different scales (for example, to evaluate a whole project, a use case, or a component). However, the more specific the tool is to the domain, the better it captures ethics readiness.

Although the tool allows for modifications, an irreplaceable part of the evaluations is the dialogue between an ethics expert and a technical expert. The dialogue is essential because it elicits ethical reflection in a way that self-assessment does not. The true value of the evaluation lies in the understanding originating from reflection and continuous exchanges. Ethical reflection develops sensitivity to the complexities of each indicator, which means engaging with the details of each question on a case-by-case basis. The tool is not meant as a rigid measurement, but as a facilitator of dialogue, because the dialogue will lead to reflection, and ultimately, to the change within the design.

**Example A: facial sketch generation**

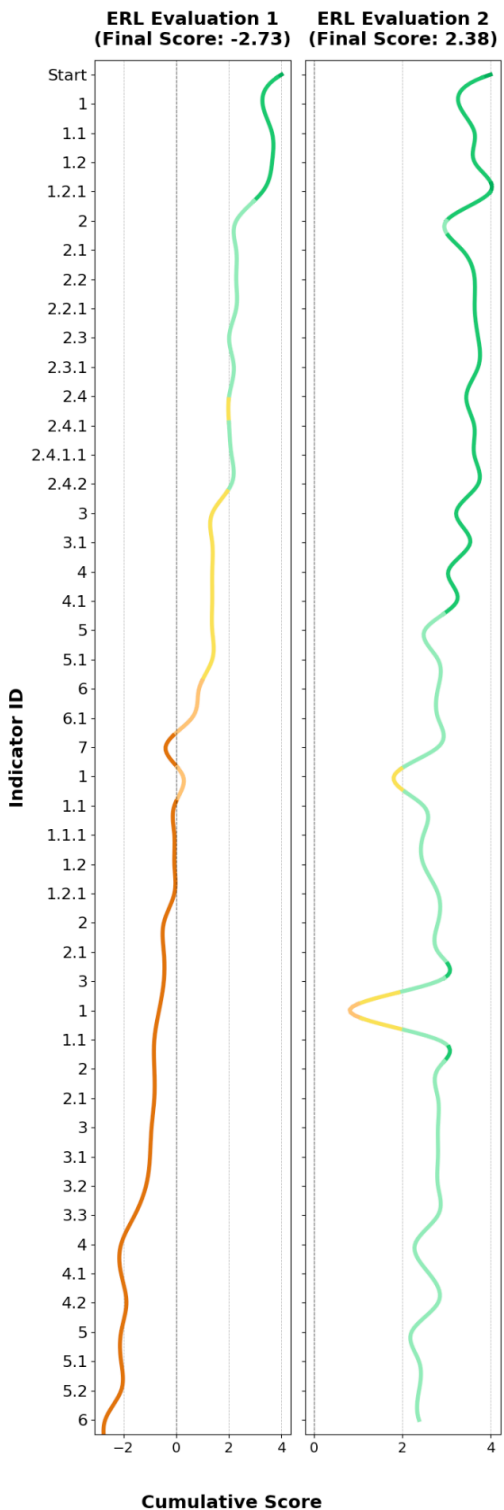

In this section we provide an example of a mock evaluation of an AI driven product for law enforcement use. The details of this mock case include a first early evaluation at TRL 4-5, and a later updated session at TRL 6-7 with a 12-month gap between the evaluation. Both sessions were attended by an ethics expert, and the product owner at the developer company, who was knowledgeable in both technology, and business aspects of the product.

The mock product being evaluated provides the generation of criminal sketches to be used by law enforcement agencies. The AI system utilizes a Generative Adversarial Network (GAN) to produce high-fidelity facial sketches from a set of descriptive parameters. The use case is simulated based on the work of Kumar et al. (Mahesh Kumar et al. 2025). The explicit purpose of the technology is to serve investigations and potentially identify suspects with a better performance than traditional manual methods involving human artists.

The first assessment was conducted when the system was a functional prototype (TRL 4-5). At this stage, the core algorithmic components are operational, but the application lacks almost any type of overall workflow and user interface. The first evaluation session takes longer than usual and lasts about 60 minutes. It is attended only by the product owner on the company's side, and an external ethics expert. During the initial discussion, and based on the nature of the product, a "Common", "AI Specific", and "Law Enforcement Specific",

"Robotics" blocks are identified. Generally, the blocks of indicators do not change between evaluation sessions.

The first evaluation takes place going through the tree-like indicator structure. Frequent stops are made to reflect on the ethical concepts and the details of the tool. Almost every answer contains a brief comment agreed by both participants. The resulting score from applying the evaluation tool was negative (less than zero). This led to a classification of the prototype at ERL 0, meaning that ethical considerations were completely lacking.

A second evaluation is performed approximately one year later, with the system having advanced to TRL 6-7. The product has become more mature following validation through pilot programs. The second assessment is conducted by the same ethics expert and the same product manager, but another senior developer is added from the product side. It is apparent that the product owner and the senior engineer have been working together on understanding and improving indicator outcomes, and have been dialogue partners with each other. Despite having more participants, the session takes less time and is finished in 40 minutes. The same blocks are selected but many more indicators point to the "validation" level, since many more "mitigation" level indicators receive a "yes" answer.

The score shows a significant improvement in the system's ethics readiness, resulting in a readiness score of 2.38. This score helped put the system at ERL 3, corresponding to a coherent and systemic analysis of ethical issues.

The feedback from the product's side includes a need for more frequent sessions, scheduling another one in 6 months instead of 12 months before the official launch of the product.

The trajectory from a negative readiness score to the score of 2.38 illustrates the power of ethics-by-design. The initial prototype represented a classic example of technological solutionism, which reframes as a quick technological fix to the complex task of drawing a face, which requires human artistry. The negative score put an emphasis on the unreflected influence of the AI system on user decision-making as well as the absence of measures to discourage over-reliance on the AI sketches. Although the system performs the technical task very well, it completely leaves out the human element, in particular the tendency of the human users to grant AI systems an authority they cannot possess (Weizenbaum 1966).

Now let us compare the two evaluation sessions. The comments from the first session indicate a factual tone:

- "No oversight processes exist,"
- "No mechanisms have been established to facilitate audits,"
- "No training has been provided."

These factual answers indicate to the product owner that they must perform design work, providing a "task" to be added to their to-do list. The overall low score in the first evaluation thus yields a relatively complete to-do list. This actionable list is presented to a colleague who also has a "problem-solving" mentality, and they form an ethics-aware duo in the company. This facilitates the progress significantly because they each have a dialogue partner to encourage reflection. First the dialogue with the ethics expert, and then internal dialogue with colleagues, helps create understanding and cycles of internal motivation to go back to the identified issues.

As we move to the second evaluation, the most visible change is that factually negative statements (e.g., "No oversight") got replaced by the rudimentary, yet real processes of accountability, auditing, and ethics oversight. This demonstrates a shift from the narrow-minded technological solutionism to reflecting upon and understanding broader societal issues, and incorporating accountability. The organisation has also created a "Dedicated reporting channel" which shows willingness to move from opaque solutionism to inviting scrutiny in a proactive way.

The second evaluation also demonstrates significant additional measures for safety, robustness, and bias mitigation, giving evidence that the designers are now aware of the principle of non-maleficence. The initial assessment revealed a neglect for the safety and security of data or for privacy ("No measures are in place to periodically review and remove unnecessary data"). While discussing these indicators with the design team, the ethics expert focused on the harms that can follow security and data breach concerns. Security issues were translated from a mere technical need, to a perspective of potentially ensuing harms to humans. In the second session, the provider showed adherence to "government/NIST frameworks" and an implementation of "Security patching and E2E encryption". During dialogue in the second session, both representatives interpreted this as measures to "protect users" rather than seeking external compliance.

The evolution of the provider's mindset between the first and the second evaluation sessions is a classic case of cognitive reframing. Initially, the product owner operated under attentional tunneling, focusing exclusively on technical function. The product's real-world risks were underestimated or neglected. The initial evaluation with its negative score and the ERL 0 result acted as a catalyst, by creating cognitive dissonance between the designer's self-perception as a mere "solution provider" and the objective harms that their technology may bring to people. The evaluation tool helped create motivation for change via creating cognitive dissonance (Festinger 1957).

The assessment process itself functions as a form of reflective gamification. By framing ethics improvement as a process of "winning back" lost points, gamifies the process and reduces tension during the evaluation. Both the ethics expert and product owner make occasional jokes about the point system. But the underlying structure of the evaluation serves as a series of Socratic prompts, creating desire to reflect on the product's flaws in anticipation of a future higher score. This iterative cycle of delayed gratification stimulates reflective design. The score, in the end, is but a symbolic representation of desire to do a better job, a symbol for the sustained and motivated effort that leads to genuine technical improvements and organisational change.

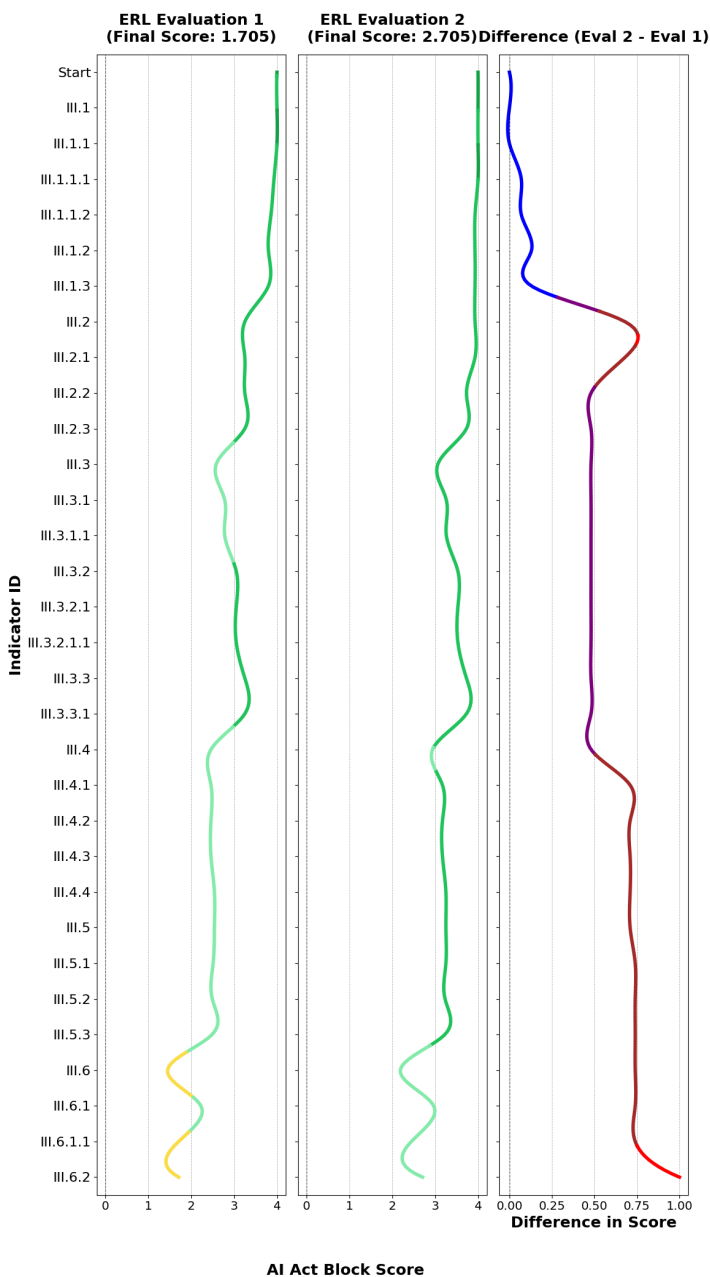

**Example B: a collaborative industrial robot**

This evaluation concerns a collaborative robot (cobot) designed for a printed circuit board (PCB) production line, where it operates in close proximity to human workers. At the soldering station, operators load freshly assembled PCBs into the cobot's secure holding tray. The cobot moves on a shared factory floor at walking speed to reach the downstream testing station, using AI-based navigation and obstacle detection components to pass safely among workers, carts, and pallets. During transit, embedded computer vision systems perform preliminary inspections for visible defects, flagging anomalies on a shared dashboard. If defects are detected, the cobot alerts the nearest available operator and reroutes the defective PCB to a rework station.

Currently at TRL 6 (prototype demonstration), the cobot is undergoing trials in a realistic manufacturing environment. Early tests have validated safe navigation in dynamic workspaces, reliable PCB transport, and in-transit defect detection. Operators already interact with the system for loading/unloading tasks, reviewing AI-generated alerts, and issuing basic rerouting commands. However, before advancing to higher TRLs, the system must undergo extended shift trials to assess endurance, robustness testing of AI vision across multiple PCB designs and lighting conditions, and scalability studies on multi-cobot coordination—including collision avoidance and route optimization.

The ERL evaluation is part of a two-phase ethics-by-design assessment framework, with a first review occurring roughly nine months before the second evaluation. At the first stage, the goal is to use ERLs to identify strengths, blind spots, and actionable improvements to maximize gains in the subsequent evaluation. Before starting the formal evaluation, the ethics expert met the development team in person. This served several purposes: familiarizing the evaluator with the operational environment, clarifying technical goals, and signaling genuine interest in the product and the team's work. Such engagement helped ensure that the assessment was not perceived as an abstract, low-priority exercise, particularly given that the evaluation process requires only a fraction of the time invested in technical development. Initial coordination between the ethics expert and the technical team proved challenging, but the informal in-person discussion established trust, facilitated smooth scheduling, and reinforced the relevance of ethics within the project.

The preliminary in-person discussion also revealed a strong initial focus on reducing repetitive strain and visual fatigue for human inspectors, alongside improving defect detection to reduce waste. Safety in human-robot interaction has been a design priority, and the AI system is explicitly positioned as supporting rather than replacing human decision-making. This reflects commendable awareness in ergonomics and safety; however, this emphasis may have overshadowed other important dimensions, such as data protection, system transparency, and user training. Data security, in particular, had not emerged as a central priority, and some vulnerabilities remained unresolved. The project manager even identified this as the area of greatest uncertainty, unable to confirm whether the AI system was resilient against adversarial attacks or model extraction attempts. Under the evaluation protocol, uncertainty is recorded as "no" rather than "I don't know," which lowers the score but also prompts internal follow-up. Notes appended to the evaluation document explicitly flagged these issues for resolution before the next round.

In this project, the evaluation process differs from a simple aggregation of all answers into a single global score. Instead, a two-step approach is used. First, a score is computed for each block, normalized on a scale from 0 to 4. These normalized block scores are then combined to produce the global score. This method was adopted because the set of blocks can vary significantly between industrial robot use cases, particularly for blocks related to GDPR and the AI Act. Computing a global score directly from all answers would penalize projects that go through more blocks, as they would have more opportunities to lose points. This could make it disproportionately harder for such projects to reach a given ERL. By first normalizing at the block level, the scoring process reduces the dependency on the number of blocks involved while preserving the integrity of the evaluation. Here, the ERL is determined as the lowest block score rather than an average. This "minimum rule" prevents one strongly performing domain from masking weaknesses elsewhere. For Project A, the AI Act block scored lowest (1.705), setting the ERL accordingly.

Crucially, the full scoring breakdown, detailed responses, and question weightings were shared with the team to support transparency and targeted improvements. This first evaluation confirmed strong performance in core areas: safety, ergonomics, and human–machine interaction. It also revealed weaker spots: assessing impacts on work organization and skills (including risks of deskilling), accounting for human diversity, improving user-facing information, and addressing security vulnerabilities.

Given the team's high engagement and openness, the identified gaps were addressed before the second evaluation, despite challenges such as technical integration issues and staff turnover. Sustaining such engagement over time is not always easy, as immediate engineering priorities can overshadow longer-term compliance efforts, making regular human contact an essential factor. In the second evaluation, several answers concerning user information were revised, leading to a marked increase in scores. The AI Act block, which had been the lowest, rose significantly from 1.705 to 2.705, while the other blocks remained relatively stable. According to the project manager, this improvement reflected a deliberate strategy: the team chose to focus on the "harder" aspects—such as organizational impact and skills—that were less natural for them, while deferring certain compliance measures, such as appointing a Data Protection Officer, to a later stage. This concentrated effort on the weakest block suggests a clear intention to raise the ERL. However, because our method defines the ERL as the lowest block score, the overall ERL remained constrained by the GDPR block at 2.300, which showed no improvement. The graph illustrates this progression, showing both the recovery of a full point within the AI Act block and the specific points at which gains were achieved. By the end of the second evaluation, although room for progress remained, the primary objective—improving AI system design through ethical deliberation—had been clearly accomplished.

**Conclusion**

Any ethical reflection, including the one about science and technology, is not an enterprise of computing a numeric score. Ethical thinking is often recurrent and cyclic, and does not proceed in a tree-like graph manner. Harmony should not be verified by algebra, to paraphrase Pushkin. Yet it is helpful to gamify and quantize ethical reflection in an industrial setting. This provides a tangible scale of progressing towards a better, more reflective and anticipatory design. It also helps engineers, who lack training in philosophy or ethics, by providing an easy-to-understand reference frame that guides their work. These benefits are real, as demonstrated by the use cases we studied in several industrial AI projects. While the use of the evaluation tool requires caution, it should be seen as an enabling instrument conducive to a win-win situation for the engineer and the ethics expert, by promoting structured dialogue between them.

With the imminent implementation of the EU AI Act and other frameworks for the responsible use and deployment of AI, a new need has emerged to develop a "standardized" interpretation of the normative principles and ethical values. Various standardization bodies and committees have been directly or indirectly tasked with creating standardized metrics, solutions, and evaluation methods that operationalize the hard normative questions and legal requirements. Even sensitive topics, such as the fair distribution of outcomes between protected groups or the monitoring of emergent AI risks, are expected to have a standardized and technical solution. However, answering such normative questions requires individual ethical reflection in context, each time giving a very specific meaning to the concepts, principles, and values in connection with the individual realities of each use case.

Technical standards act more like a generic frame or a broad "test" performed at the end of the development cycle to check conformity of the final product. Ethics by design, on the other hand, is supposed to take place during the development to have a cumulative and continuous effect on the technical process while it is not too late to influence or modify it. This ongoing, reflective and anticipatory approach should not be outsourced to a generic compliance framework, because the overall goal is to create intrinsic and efficient motivation for the engineers, leading to technical improvements that are embedded into a system. The ERL methodology achieves this ongoing engagement through human dialogue, not only numeric scores. The ERL evaluation is designed to help reflect on the reasons, causes, and consequences of the choices made by the engineer, and to make them aware of the underlying moral and societal issues. A technical *post hoc* standard is not fit for this goal, for it remains, by definition, an external rule-based approach.

The focus on standards and norms reduces AI ethics to the narrow deontological dimension of duty and blame. It frames ethical reflection as merely avoiding sanction by adhering to rules. Yet, the engineer's responsibility is broader and his moral stance much richer (Alexei Grinbaum and Groves 2013). As it includes hard-to-quantify concepts like user trust and desire, technology attraction and seduction, empathy in human-machine interaction, or moral luck, where outcomes are beyond an agent's control and judgment depends on future random events, the only form of human thinking that can help the engineer grasp this complex reality is dialogical. Yet any dialogue - and a dialogue about technology in particular - requires structure to achieve efficiency. The ERL scale and evaluation tool provide just that.

**Data**

Our developed and validated ERL indicators are divided into 5 blocks:

1) Common block
2) Privacy block
3) Law enforcement block
4) AI block
5) Robotics block

They are available on github under GNU General Public License v3.0:
https://github.com/LA-NS/ethics-readiness-levels

**References**

Adomaitis, Laurynas, Alexei Grinbaum, and Dominic Lenzi. 2022. "TechEthos D2. 2: Identification and Specification of Potential Ethical Issues and Impacts and Analysis of Ethical Issues of Digital Extended Reality, Neurotechnologies, and Climate Engineering." PhD Thesis. CEA Paris Saclay. https://cea.hal.science/cea-03710862/ (May 29, 2024).

Adomaitis, Laurynas, Björn Hoog, and Alexei Grinbaum. 2024. "Security and Ethics Readiness Levels: Two New Scales." *2024 IEEE International Conference on Technology Management, Operations and Decisions (ICTMOD)*: 1–8. doi:10.1109/ICTMOD63116.2024.10878193.

Bietti, Elettra. 2021. "From Ethics Washing to Ethics Bashing: A View on Tech Ethics from within Moral Philosophy." In SSRN. https://papers.ssrn.com/sol3/papers.cfm?abstract_id=3914119.

Brey, Philip. 2010. "Values in Technology and Disclosive Computer Ethics." *The Cambridge handbook of information and computer ethics* 4: 41–58.

Brey, Philip, and Brandt Dainow. 2021. "Ethics by Design and Ethics of Use in AI and Robotics." *The SIENNA project-Stakeholder-informed ethics for new technologies with high socio-economic and human rights impact. Accessed April* 26: 2021.

Buber, Martin. 2023. *I and Thou*. New York: Scribner.

Calo, Ryan. 2015. "Robotics and the Lessons of Cyberlaw." *California Law Review* 103(3): 513–63.

Dupuy, Jean-Pierre, and Alexei Grinbaum. 2005. "Living with Uncertainty: From the Precutionary Principle to the Methodology of Ongoing Normative Assessment." *Geoscience* 337: 457–74.

Eljasik-Swoboda, Tobias, Christian Rathgeber, and Rainer Hasenauer. 2019. "Assessing Technology Readiness for Artificial Intelligence and Machine Learning Based Innovations." In *DATA*, , 281–88.

European Commission. 2020. *The Assessment List for Trustworthy Artificial Intelligence (ALTAI)*. European Commission.

European Commission. 2022. *Proposal for a REGULATION OF THE EUROPEAN PARLIAMENT AND OF THE COUNCIL LAYING DOWN HARMONISED RULES ON ARTIFICIAL INTELLIGENCE (ARTIFICIAL INTELLIGENCE ACT) AND AMENDING CERTAIN UNION LEGISLATIVE ACTS*. https://eur-lex.europa.eu/legal-content/EN/TXT/?uri=CELEX%3A52021PC0206 (March 24, 2022).

Festinger, Leon. 1957. *A Theory of Cognitive Dissonance*. Stanford University Press.

Floridi, Luciano. 2016. "Faultless Responsibility: On the Nature and Allocation of Moral Responsibility for Distributed Moral Actions." *Philosophical Transactions of the Royal Society A: Mathematical, Physical and Engineering Sciences* 374(2083): 20160112. doi:10.1098/rsta.2016.0112.

Grinbaum, A. 2019. *Les robots et le mal*. Desclée de Brouwer.

Grinbaum, Alexei, Raja Chatila, Laurence Devillers, Jean-Gabriel Ganascia, Catherine Tessier, and Max Dauchet. 2017. "Ethics in Robotics Research: CERNA Mission and Context." *IEEE Robotics & Automation Magazine* 24(3): 139–45. doi:10.1109/MRA.2016.2611586.

Grinbaum, Alexei, and Christopher Groves. 2013. "What Is 'Responsible' about Responsible Innovation? Understanding the Ethical Issues." *Responsible innovation: Managing the responsible emergence of science and innovation in society*: 119–42.

Guston, David H. 2014. "Understanding 'Anticipatory Governance.'" *Social Studies of Science* 44(2): 218–42. doi:10.1177/0306312713508669.

Hagendorff, Thilo. 2020. "The Ethics of AI Ethics: An Evaluation of Guidelines." *Minds and Machines* 30(1): 99–120. doi:10.1007/s11023-020-09517-8.

de Hert, Paul, and Vagelis Papakonstantinou. 2016. "The New Police and Criminal Justice Data Protection Directive: A First Analysis." *New Journal of European Criminal Law* 7(1): 7–19. doi:10.1177/203228441600700102.

Jonas, Hans. 1985. *The Imperative of Responsibility: In Search of an Ethics for the Technological Age*. University of Chicago Press.

Jong, Eline de. 2025. "Ethics Readiness of Technology: The Case for Aligning Ethical Approaches with Technological Maturity." doi:10.48550/arXiv.2504.03336.

Kratsios, Michael J., David O. Sacks, and Marco A. Rubio. 2025. *America's AI Action Plan*. Executive Office of the President of the United States.

Kundu, Sandipan, Yuntao Bai, Saurav Kadavath, Amanda Askell, Andrew Callahan, Anna Chen, Anna Goldie, et al. 2023. "Specific versus General Principles for Constitutional AI." doi:10.48550/arXiv.2310.13798.

Lavin, Alexander, Ciarán M. Gilligan-Lee, Alessya Visnjic, Siddha Ganju, Dava Newman, Sujoy Ganguly, Danny Lange, et al. 2022. "Technology Readiness Levels for Machine Learning Systems." *Nature Communications* 13(1): 6039. doi:10.1038/s41467-022-33128-9.

Lessig, Lawrence. 2000. "Code Is Law." *Harvard magazine* 1: 2000.

Levinas, Emmanuel. 1991. *Totality and Infinity*. Dordrecht: Springer Netherlands. doi:10.1007/978-94-009-9342-6.

Mahesh Kumar, S., R. Elangovan, K. R. Goutham, and S. Jayamani. 2025. "AI-Powered Face Sketching for Criminal Identification." In *Intelligent Computing and Communication*, eds. K. Srujan Raju, Roman Senkerik, T. Kishore Kumar, Mathini Sellathurai, and Voruganti Naresh Kumar. Singapore: Springer Nature, 415–25. doi:10.1007/978-981-96-1267-3_35.

Moor, James H. 1985. "What Is Computer Ethics?" *Metaphilosophy* 16(4): 266–75. doi:10.1111/j.1467-9973.1985.tb00173.x.

Olechowski, Alison, Steven D. Eppinger, and Nitin Joglekar. 2015. "Technology Readiness Levels at 40: A Study of State-of-the-Art Use, Challenges, and Opportunities." *2015 Portland International Conference on Management of Engineering and Technology (PICMET)*: 2084–94. doi:10.1109/PICMET.2015.7273196.

Politi, Vincenzo, and Alexei Grinbaum. 2020. "The Distribution of Ethical Labor in the Scientific Community." *Journal of Responsible Innovation* 7(3): 263–79.

Richardson, Henry S. 1990. "Specifying Norms as a Way to Resolve Concrete Ethical Problems." *Philosophy & Public Affairs* 19(4): 279–310.

Sauser, Brian, Dinesh Verma, Jose Ramirez-Marquez, and Ryan Gove. 2006. "From TRL to SRL: The Concept of Systems Readiness Levels." In *Conference on Systems Engineering Research, Los Angeles, CA*, Citeseer, 1–10.

Stahl, Bernd Carsten, Job Timmermans, and Catherine Flick. 2017. "Ethics of Emerging Information and Communication TechnologiesOn the Implementation of Responsible Research and Innovation." *Science and Public Policy* 44(3): 369–81.

Tang, Yong, Jason Xiong, Rafael Becerril-Arreola, and Lakshmi Iyer. 2020. "Ethics of Blockchain." *Information Technology & People* 33(2): 602–32. doi:10.1108/ITP-10-2018-0491.

Terpan, Fabien. 2015. "Soft Law in the European Union—The Changing Nature of EU Law." *European Law Journal* 21(1): 68–96. doi:10.1111/eulj.12090.

Umbrello, Steven, Michael J. Bernstein, Pieter E. Vermaas, Anaïs Resseguier, Gustavo Gonzalez, Andrea Porcari, Alexei Grinbaum, and Laurynas Adomaitis. 2023. "From Speculation to Reality: Enhancing Anticipatory Ethics for Emerging Technologies (ATE) in Practice." *Technology in Society* 74: 102325.

Van den Hoven, Jeroen, Pieter E. Vermaas, and Ibo Van de Poel. 2015. *Handbook of Ethics, Values, and Technological Design: Sources, Theory, Values and Application Domains*. Springer.

Von Schomberg, René. 2013. "A Vision of Responsible Research and Innovation." In *Responsible Innovation*, eds. Richard Owen, John Bessant, and Maggy Heintz. Wiley, 51–74. doi:10.1002/9781118551424.ch3.

Weizenbaum, J. 1966. "ELIZA - A Computer Program for the Study of Natural Language Communication between Man and Machine"." *Communications of the Association for Computing Machinery* 9: 36–45.

Williams, Bernard. 2006. *Ethics and the Limits of Philosophy*. Routledge.